# Robust control design for multi-input multi-output plasma shape control on EAST tokamak using H∞ synthesis

Lei Liu, Bingjia Xiao, Yong Guo, Yuehang Wang

*Abstract*—Accurate plasma shape control is the basis of tokamak plasma experiments and physical research. Modeling of the linearized control response of plasma shape and position has been widely used for shape controller design in the last several years. But it usually contains much of the uncertainty, such as structured uncertainties and unmodeled dynamics. EAST tokamak plasma shape controller design is also based on a linear rigid plasma response model which integrated within a Matlab-based toolset known as TokSys. Meanwhile the PID control approach is currently used for EAST plasma shape control. This leads to strong coupling between different parameters describing the plasma shape. To handle these problems, a H∞ robust control scheme for EAST multi-input multi-output (MIMO) shape control has been proposed. First, the plasma response is modeled as the linearized rigid RZIp model. Then, the controller design technique is introduced with two main stages: 1) loop shaping is used to shape the nominal plant singular values to give desired open-loop properties at frequencies of high and low loop gain; 2) a normalized coprime factorization and H∞ technique is used to decouple the most relevant control channels and minimize the tracking errors. Finally, the simulation results show that the H∞ robust controller combines good robust stability margins, speed of response, dynamic tracking characteristics, and closed-loop decoupling for EAST plasma shape control.

H∞ robust control, EAST, rigid plasma model, plasma shape decoupling control, MIMO

## I. INTRODUCTION

In modern tokamaks, plasma are elongated to occupy as much volume as possible for energetic reasons, this means that plasma must be maintained as close as possible to nearby components such as the first wall. To achieve this and maintain good stable performance, including the desired plasma current, the desired plasma shape and position, an accurate and robust shape control must be guaranteed. The main objective of plasma shape control in tokamak is achieved by a set of toroidal coils wrapped around the vacuum vessel, called Poloidal Field (PF) coils. The current, position and shape control has been subject to study and research in last decade [1-3]. The objective of these is to track desired trajectories of the controlled variables, which maintaining the plasma at a desired equilibrium point during the flat-top phase. In most of the previous work, the simplest single input sing output (SISO) PID-based controller is widely used. And in general, the controller has to be refined by experiential tuning. Then later, multivariable plasma magnetic controllers have been investigated in the most advanced tokamaks all around the world, such as JET [4 ,5], DIII-D [6], TCV [7], NSTX [8].

This paper describes a model-based multi-input multi-output (MIMO) algorithm for plasma shape and position of EAST tokamak using H∞ synthesis. The design of EAST plasma shape controller based on a linear rigid plasma response model (RZIP model) which integrated within a Matlab-based toolset known as TokSys [9]. Meanwhile the single input single output (SISO) PID-based controller is routinely used for EAST plasma shape control, which is briefly introduced in section II. This SISO control leads to strong coupling between different parameters describing the plasma shape. To handle these problems, a H∞ robust control scheme for EAST MIMO shape control has been proposed. First, the RZIp model, which describes the measured and controlled variables in terms of the state variables, has been built. This is described in section III. Then, the controller design technique is introduced with two main stages in section IV: 1) loop shaping is used to shape the nominal plant singular values to give desired open-loop properties at frequencies of high and low loop gain; 2) a normalized coprime factorization and H∞ technique is used to decouple the most relevant control channels and minimize the tracking errors. In section V, the preliminary simulation results of EAST shape control are presented using this method. Finally, the conclusion is given in section VI.

## II. EAST PLASMA SHAPE ISOFLUX CONTROL SCHEMES

Plasma magnetic control on EAST is achieved by driving the required currents in the poloidal field (PF) coil system. Fig.1 shows the poloidal cross-section of the EAST tokamak. There are 14 superconductive coils (PF1-14) (the couples PF7/PF9 and PF8/PF10 are connected in series), and 2 in-vessel copper coils (IC1 and IC2). IC1 and IC 2 are connected in anti-series in order to control plasma vertical instability on a faster time scale comparing with the ex-vessel PF coils.

EAST Plasma Control System (PCS) is in charge of controlling the current in the PF coils system [10], [11], which adopted from the DIII-D PCS [12]. Fig.2 shows the simplified block diagram of EAST PCS. It's magnetic control system mainly including the following control logic:
.

October 31, 2020. His work was supported by the National Natural Science Foundation of China under Grant No.11805235, and National MCF Energy R&D Program (Grant No. 2018YFE0302100, No. 2018YFE0302101).

Lei Liu, is with Institute of Plasma Physics, Chinese Academy of Sciences, Hefei, China. (e-mial:liulei@ipp.ac.cn).

- plasma current control
- shape and position control
- vertical instability control

In paper [11,13], control logic for plasma current, basic position control, and shape control in EAST PCS are described in detail. This paper only discus the plasma shape and position control which deals with the control of the position of the plasma centroid, and of the shape of the plasma boundary. For EAST plasma shape and position control, two algorithms are currently available which is RZIP control and isoflux control. RZIP control only control the radial and vertical position of plasma centroid, usually PF11-PF14 coils are used to control the centroid position while other PF circuits are used only for plasma current control. RZIP is to simple too be used for accurate plasma shape control. Isoflux control is mainly used for EAST plasma shape control, that aims at controlling the shape of the plasma boundary, by controlling the position of the X-point and by regulating to zero the flux error on a set of control points. Fig .3 shows the plasma target shape (black) and the control segment line(red). The control points are the intersection between the plasma boundary and the control segments. For both the plasma current and plasma shape and position control, proportional-integral-derivative (PID) controllers are routinely used in EAST experiment campaign. For each control channel, the PID outputs are distributed over the 12 available superconductive circuits, by using the weights specified into the correspondent column of the so-called M matrix (in Fig.2). But the M matrix is a sparse matrix, that means each controlled variable is linked to a subset of PF circuits. It is obviously not a MIMO controller but a SISO controller. This SISO controller works well for most discharge scenarios until now, but works weakly for complex plasma shape such as snow flake equilibria on EAST tokamak. Although a MIMO architecture for plasma shape [14,15] has been design and tried on EAST, but which still using PID+M-matrix control logic. The mainly difference is the M matrix is a full rank matrix and all PF coils are used for each segment control comparing to the routinely controller. Although it is a MIMO controller, but the control robust issue has not been considered. In this paper, we aim to design a more robust plasma shape controller, which able to work with different scenarios, independently of the desired plasma shape.

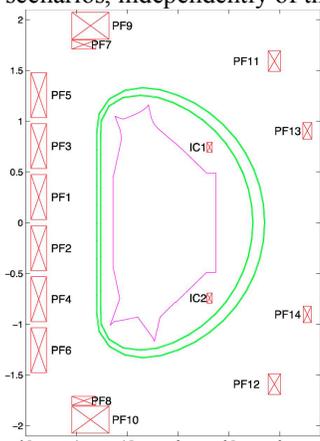
Fig. 1. EAST geometry of Poloidal Field (PF),the double layer vacuum vessels, the in-vessel coils(IC).

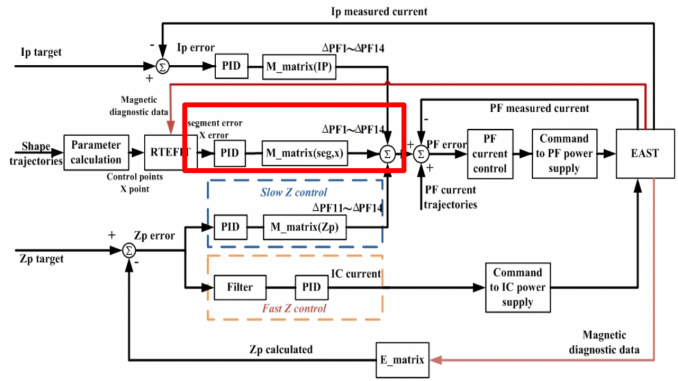
Fig. 2. Isoflux control scheme of EAST PCS.

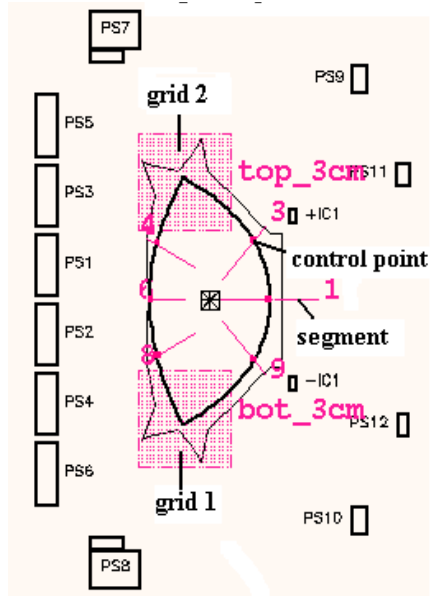
Fig. 3. EAST plasma target shape (black) and the control segment line(red).

The Introduction motivates the work. Explain here why it is important, what deficiencies in the state of the art you aim to overcome, how/why the work extends the state of the art. Describe the problem you are trying to solve. Include a review of relevant literature in this section.

In the Experimental section, you describe your equipment, technique, materials, protocols, and whatever else is necessary so that another person working in your field can understand what you did and reproduce the work.

Discuss the data you acquired and how you analyzed them. Include a statistical analysis to demonstrate the limits of the data analysis. Experimental data have uncertainties, although the error bars may be smaller than the symbols used to represent them. Your discussion should include something about the experimental uncertainties. Even simulations have (or can have) uncertainties associated with them. TNS expects to see a discussion of uncertainties in all papers.

A Conclusions (note the plural) section is required. Although this section may review the main points of the paper, do not replicate the abstract here. A Conclusions section draws inferences from the data and elaborates on the importance of the work. You may suggest applications and extensions. You should explain here how the data and analysis support the introduction to your paper where you motivated and justified the work. Explain how the work extends the state of the art. Answer the question "So what?"

## III. LINEAR RIGID PLASMA RESPONSE MODEL OF EAST

After you open TNS_Word_Template.docx, type over sections or cut and paste from another document while retaining the styles in this template. The various sections and headings in this document are formatted according to styles that can be found *Word*'s Styles menu. For example, this section uses the "Text" style. As you overwrite this document, retain the existing Word style. Do not change the font sizes or line spacing to squeeze more text into a limited number of pages. *Use italics for emphasis*; do not underline.

A tokamak device is a rather complex system, including the plasma, the active coils, and the metallic structures. We are mainly interested in the electromagnetic interaction of the plasma with the surrounding coils and the control of the plasma vertical position. For these purposes, the linearized RZIP model which is verified by DIII-D [9] and EAST [16] are used.

In the model [9], the toroidal symmetry is assumed. It is also assumed that only toroidal currents flowing in the plasma, active coils and surrounding passive structures. Plasma current is represented by a fixed spatial distribution of current from particular plasma equilibrium and only rigid radial and vertical motion is allowed. The poloidal field coil system must be driven by power supplies, while no any external voltages are applied to the passive structure. Vacuum vessel is discretized into a number of toroidal elements each of which carries an individual toroidal current.

The system of plasma, shaping coils, and passive structure can be described using circuit equations derived from Faraday's Law. The circuit model representation of the external toroidal conductors and plasma circuit can be expressed as [9]:

$$\frac{dM_{ss}I_s}{dt} + R_{ss}I_s + M_{sp}\frac{dI_p}{dt} + I_{p_0}\frac{\partial M_{sp}}{\partial R_c}\frac{\partial R_c}{\partial t} + I_{p_0}\frac{\partial M_{sp}}{\partial Z_c}\frac{\partial Z_c}{\partial t} = V_s \quad (1)$$

$$L_p\frac{dI_p}{dt} + R_pI_p + M_{ps}\frac{dI_s}{dt} + I_{s_0}\frac{\partial M_{ps}}{\partial R_c}\frac{\partial R_c}{\partial t} + I_{s_0}\frac{\partial M_{ps}}{\partial Z_c}\frac{\partial Z_c}{\partial t} = 0 \quad (2)$$

The model can be expressed in state space form:

$$\frac{d}{dt}\begin{bmatrix}I_s\\I_p\end{bmatrix} = A\begin{bmatrix}I_s\\I_p\end{bmatrix} + B\begin{bmatrix}V_s\\0\end{bmatrix} \quad (3)$$

Where,

$$A = -M^{-1}\begin{bmatrix}R_{ss} & 0\\0 & R_p\end{bmatrix} \quad (4)$$

$$B = M^{-1} \quad (5)$$

$$M = \begin{bmatrix}M_{ss}+X_{ss} & M_{sp}+X_{sp}\\M_{ps}+X_{ps} & L_p+X_{pp}\end{bmatrix}, \quad (6)$$

$$X_{ss} = I_{p_0}\frac{\partial M_{sp}}{\partial R_c}\frac{\partial R_c}{\partial I_s} + I_{p_0}\frac{\partial M_{sp}}{\partial Z_c}\frac{\partial Z_c}{\partial t}, \quad (7)$$

$$X_{sp} = I_{p_0}\frac{\partial M_{sp}}{\partial R_c}\frac{\partial R_c}{\partial I_p} + I_{p_0}\frac{\partial M_{sp}}{\partial Z_c}\frac{\partial Z_c}{\partial I_p}, \quad (8)$$

$$X_{pp} = I_{p_0}\frac{\partial M_{ps}}{\partial R_c}\frac{\partial R_c}{\partial I_p} + I_{p_0}\frac{\partial M_{ps}}{\partial Z_c}\frac{\partial Z_c}{\partial I_p}, \quad (9)$$

$$X_{ps} = I_{p_0}\frac{\partial M_{ps}}{\partial R_c}\frac{\partial R_c}{\partial I_s} + I_{s_0}\frac{\partial M_{ps}}{\partial Z_c}\frac{\partial Z_c}{\partial I_s}, \quad (10)$$

Here, subscript"s" refers to all stabilizing conductors and"p" refers to plasma. Where $M_{ss}$ and $R_{ss}$ are the conductor-to-conductor mutual inductance matrix and the conductor resistance matrix, $I_s$ is the conductor current vectors, $I_p$ is the plasma grid current vectors, $X_{ss}$ the represents the variation of conductor flux due to plasma motion (the change of plasma radial and vertical position of the current centroid) in response to conductor current variation, $X_{sp}$ represents variation in conductor flux due to plasma motion in response to plasma current variation, $M_{sp}$ is the mutual coupling from plasma to conductors, $V_s$ is the vector of voltages of all the PF coils. $X_{ps}$ represents variation in plasma flux due to plasma motion in response to conductor current variation, $M_{ps}$ is the mutual coupling from conductors to plasma, $X_{pp}$ represents the variation of plasma flux due to plasma motion in response to plasma current variation, $L_p$ is the plasma self-inductance.

The tokamak outputs like the plasma parameters and the magnetic diagnostic measurements are linear combinations of the state variables. An output equation describes the measured and controlled variables (R&Z position of plasma, perturbed plasma current, perturbed fluxes at the isoflux points, perturbed fields at X-point grid positions) in terms of the state variables (Is and Ip) can be expressed as:

$$\begin{pmatrix}\delta R\\\delta Z\\\delta I_p\\\delta \psi\\\delta B_{grid}\end{pmatrix} = \begin{pmatrix}\frac{\partial R}{\partial I_s} & \frac{\partial R}{\partial I_p}\\\frac{\partial Z}{\partial I_s} & \frac{\partial Z}{\partial I_p}\\\frac{\partial I_p}{\partial R}\frac{\partial R}{\partial I_s} & \frac{\partial I_p}{\partial Z}\frac{\partial Z}{\partial I_s}\\M_{sg}+I_p\frac{\partial M_{sg}}{\partial Z}\frac{\partial Z}{\partial I_s}+I_p\frac{\partial M_{sg}}{\partial R}\frac{\partial R}{\partial I_s} & M_{pg}+I_p\frac{\partial M_{pg}}{\partial Z}\frac{\partial Z}{\partial I_p}+I_p\frac{\partial M_{pg}}{\partial R}\frac{\partial R}{\partial I_p}\\\frac{\partial B}{\partial I_s}+\frac{\partial B}{\partial R}\frac{\partial R}{\partial I_s}+\frac{\partial B}{\partial Z}\frac{\partial Z}{\partial I_s} & \frac{\partial B}{\partial I_p}+\frac{\partial B}{\partial R}\frac{\partial R}{\partial I_p}+\frac{\partial B}{\partial Z}\frac{\partial Z}{\partial I_p}\end{pmatrix}\begin{pmatrix}\delta I_s\\\delta I_p\end{pmatrix}$$

, (11)

Expression of the plasma-conductor system in this linear state form allows design of multivariable controllers with explicit optimization of various performance and robustness characteristics. Equation (3) and (11) can be expressed in state space form,

$$\dot{x} = Ax + Bu, \quad (12)$$
$$y = Cx + Du, \quad (13)$$

with

A = System matrix
B = Control input matrix
x = System states vector

u = Input vector
C = Output matrix
D = Direct transmission matrix
y = output vector

A matrix contains the system dynamics, The B matrix is used to align the input with the system.The D matrix is the direct transmission matrix of input to output. For most cases this is the null matrix, because there usually is no such connection.

## IV. H∞ CONTROL SYSTEM DESIGN

In ref. [7], the robust controller design method using the H∞ normalized comprime factorization has been discussed in detail. It gives the central controller in the form of a right comprime factorization:

$$K_{11} = \begin{bmatrix} A - BB'X - \beta^{-2}ZYC'C & ZYC' \\ -\beta^{-2}B'X & 0 \end{bmatrix}, \quad (14)$$

Where A,B,C are the state space expression of the RZIp model built in section 3. $X, Y, \beta, \gamma, Z$ can be calculated by according the all suboptimal controllers design method described in ref. [7] which not be list here again.

The H∞ controller $K_{11}$ decoupled the EAST plasma shape control while it can also maximum the robustness of the system. Fig.4 shows the control logic of shape and position control using H∞ controller. Here, the controller is specified by $K_{11}$, the "Tokamak Model" is the RZIp response model described in section 3. The "PF Power supply" is the 14 PF coils which it actuated by 12 individual power supply.

For EAST PF power supply, it is modeled as a first order lowpass filter with 3.3ms response delay time (communication delay and power supply response time). Of course, the PF coils power supply capability also has its limitation. In this paper, we only consider the voltage limitation of power supply. The current rising rate limitation of the superconductor PF coils are ignored, which to protect the superconductive coils from overheating, because it usually within the limitation. Table 2 lists the voltage limitation of EAST PF coil power supply,which also modeled in the "PF power supply" model.

TABLE I
VOLTAGE LIMITATION OF POWER SUPPLY EAST PF COILS

| PF coils power supply limitation | Maximum output (V) |
|---|---|
| PF1-PF6 | 350 |
| PF7/PF8 | 1110 |
| PF9/PF10 | 700 |
| PF11/PF12 | 350 |

In Fig.4, W1 and W2 is used for loop shaping. In general, the weight functions W1 and W2 are chosen arbitrarily, and then tuned manually to meet the desired properties of controller. In this paper, the simplest weight functions are used, which have one pole and zero. W1 and W2 are used to shape the nominal plant singular values to give desired open and closed loop properties at frequencies of high and low loop gain. The H∞ controller characteristics is achieved by tuning the parameters of weight functions. Fig.5 shows the singular values of unweighted EAST RZIp model (which including the PF power supply model here) with 12 inputs and 9 outputs. It can be seen from the figure that it has a large spread in singular value magnitudes. This will lead to highly "direction sensitive" responses, it was necessary to rescale the outputs, thereby balancing the singular values. After weighted by W1 and W2, Fig. 6 show that the singular values of weighted closed-loop transfer functions have much smaller DC gains. This means that the decoupling of the control outputs be insured.

To sum up, with the entire set of the synthesized controllers, we desire the best decoupling of all the relevant control channels and also minimize the tracing errors.

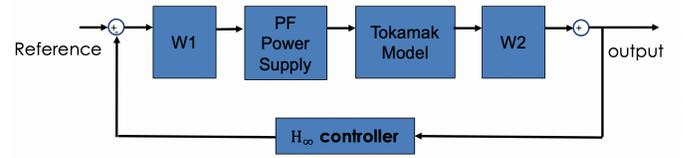

Fig. 4. Control logic of EAST plasma shape control using H∞ control synthesized

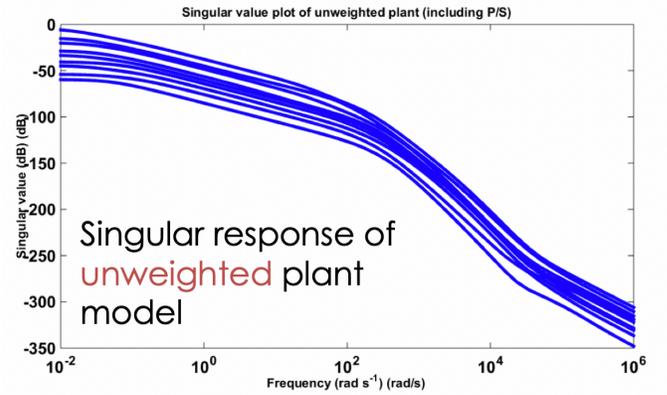

Fig. 5. Singular response of unweighted plant model

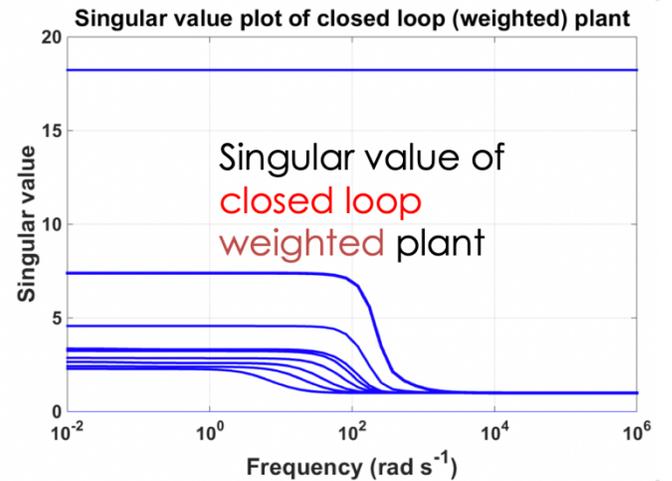

Fig. 6. Singular response of unweighted plant model

## V. SIMULATION RESULTS

Six control segments named seg01/03/04/06/09 and the radial and vertical position of X-point in grid 1 area showed in Fig.2 are selected to define the plasma shape and the magnetic flux on the boundary. The control points are the intersection between the plasma boundary and the control segments. It is necessary to control the poloidal flux at the six control points and also the flux at the X-point, the radial and vertical position of X-point should be controlled to the targets meanwhile.

In order to test the robust stability margins, speed of response, dynamic tracking characteristics and closed loop decoupling. The pulse amplitudes are chosen to test the system responses and the inter-loop cross coupling. Fig.7 shows the performance of the closed loop feedback control. Red line is the reference, blue line is the simulated responses. It is clear show that good decoupling of each control segment, only cross coupling appears at the pulse edges. The controller possesses all the desired properties. It shows good tracking properties such as speed of response and setting time.

Fig. 8 shows PF coil voltage requirements for all 12 PF power supply. Some of them are saturated because the voltage limits during the pulse edge. All PF power supply are actuated for every segment control. Good decoupling of plasma shape control and response tracking even with the power supply saturation.

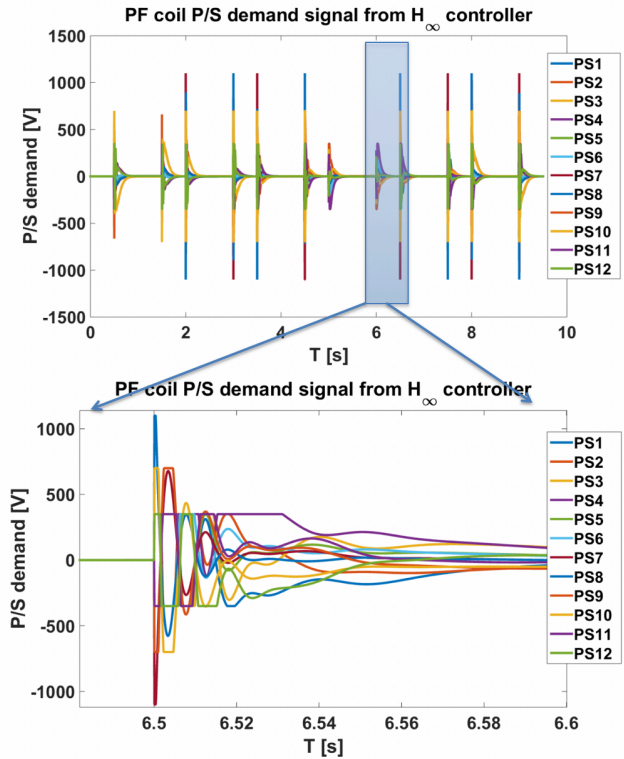

Fig. 8.  Differential (a) and integral probability of interaction as a function of angle of incidence. Note that "Fig." is abbreviated. There is a period after the figure number, followed by two spaces. PF coils voltage requirements for all 12 power supply.

## VI. CONCLUSION

A new MIMO controller for EAST plasma shape isoflux control has been preliminary developed. Beginning with the design, the linearized rigid RZIp model of EAST tokamak has been built. Then using loop shaping, a normalized coprime factorization and H∞ synthesis, the robust H∞ controller has been built. Finally, the simulation results show that the H∞ robust controller combines good robust stability margins, speed of response, dynamic tracking characteristics, and closed-loop decoupling for EAST plasma shape control. Decoupling control and response tracking guaranteed even with the power supply voltage saturation limits.

ACKNOWLEDGMENT

This work was supported by the National Natural Science Foundation of China under Grant No.11805235, and National MCF Energy R&D Program (Grant No. 2018YFE0302100, No. 2018YFE0302101).

REFERENCES

*Basic format for books:*
 J. K. Author, "Title of chapter in the book," in *Title of His Published Book,* x*th ed. City of Publisher, (only U.S. State), Country: Abbrev. of Publisher, year, ch. x, sec. x, pp. xxx–xxx.*
*Examples:*
[1] Ariola M. and Pironti A. 2008 Magnetic control of tokamak plasmas (Berlin: Springer)
[2] De Tommasi G. et al 2011 Fusion Sci. Technol.59 486–98
[3] Ambrosino R. et al 2015 IEEE Conference on Control Applications (Sydney, Australia) pp 1290–5
[4] G. De Tommasi et al., Shape control with the eXtreme Shape Controller during plasma current ramp-up and ramp-down at JET tokamak, J. Fusion Energ., 33(2014) 233–242
[5] G. Ambrosino, M. Ariola, A. Pironti, F. Sartori, Design and implementation of an output regulation controller for the JET tokamak. IEEE Trans. Control Syst. Technol. 16(6), 1101–1111
[6] Humphreys D.A., Walker M.L.,et al, Initial implementation of a multivariable plasma shape and position controller on the DIII-D tokamak. In Proceeding of the 2000 IEEE International Conference on Control Applications, 2000
[7] Sharma A.S., et al., 2002, Tokamal modeling and control. PhD thesis
[8] W. Shi, M. Alsarheed, E. Schuster, M.L. Walker, J. Leuer, D.A. Humphreys, D.A. Gates, Fusion Eng. Des. 86 (2011) 1107–1111
[9] Humphreys D.A., Walker M.L., Welander A.S., Minimal plasma response models for design of tokamak equilibrium controllers with high dynamic accuracy. Proc. 41st Annu. Meeting of Division of Plasma Physics, Seattle, WA.
[10] B. J. Xiao et al. Enhancement of EAST plasma control capabilities. Fus. Eng. Des., 112:660–666, 2016.
[11] Q. P. Yuan et al. Plasma current, position and shape feedback control on EAST. Nucl. Fus., 53(4):043009, Apr. 2013.
[12] Humphreys D. et al 2008 Fusion Eng. Des. 83 193
[13] Tommasi, G. De , et al. On plasma vertical stabilization at EAST tokamak. 2017 IEEE Conference on Control Technology and Applications (CCTA) IEEE, 2017.
[14] Guo, Y. , et al. Preliminary results of a new MIMO plasma shape controller for EAST. Fusion Engineering & Design 128.MAR.(2018):38-46.
[15] Albanese, R. , et al. A MIMO architecture for integrated control of plasma shape and flux expansion for the EAST tokamak. Control Applications IEEE, 2016.
[16] L. Liu, B.J. Xiao, D.A. Humphreys et al, Fusion Engineering and Design 89, 563-567 (2014)